\documentclass{article}
\usepackage{graphicx}
\usepackage[round]{natbib}
\usepackage{epsfig}
\title{
Predicting hurricane numbers from Sea Surface Temperature: closed
form expressions for the mean, variance and standard error of the
number of hurricanes }

\author{Stephen Jewson (RMS)\footnote{\emph{Correspondence email}: \texttt{stephen.jewson@rms.com}}\\}

\begin{document}
\maketitle

\begin{abstract}
One way to predict hurricane numbers would be to predict sea surface temperature, and then
predict hurricane numbers as a function of the predicted sea surface temperature. For certain
parametric models for sea surface temperature and the relationship between sea surface temperature
and hurricane numbers, closed-form solutions exist for the mean and the variance of the number
of predicted hurricanes, and for the standard error on the mean. We derive a number of such
expressions.
\end{abstract}

\section{Introduction}

One way to try and predict future hurricane numbers is to predict sea surface temperatures (SST),
and then to predict hurricane numbers as a function of the predicted SSTs.  If both the
prediction of SSTs and the model that relates SSTs
to hurricane numbers are probabilistic then the resulting probabilistic prediction of hurricane numbers can,
in general, only be derived using numerical methods. However, in certain cases a lot of information about the predicted distribution of hurricane numbers
can be derived analytically. This includes estimates for the mean number of hurricanes,
the variance of the number of hurricanes,
the standard error on the estimate of the mean number of hurricanes and
the linear sensitivity of the mean number of hurricanes to changes in the mean and variance of SST.

In this article we derive a number of such relations, for the following
two cases:
\begin{enumerate}

    \item we predict the SST distribution, and then use these predicted SSTs
     to predict either basin hurricane numbers or landfalling hurricane numbers.

    \item we predict the SST distribution, use this to predict the distribution
    of basin hurricane numbers, and then use a further relationship to predict landfalling hurricane
    numbers from basin hurricane numbers.

\end{enumerate}

The assumptions we make to render this problem tractable are as follows:

\begin{itemize}

    \item the sea surface temperature is taken as normally distributed
    (although in several of our derivations we relax this assumption and consider
    a completely general distribution of SST with known mean and variance)

    \item the relationship between sea surface temperature and hurricane numbers
    is taken as either (a) linear and normally distributed, (b) linear and poisson distributed,
    or (c) exponential and poisson distributed.

    \item the relationship between hurricane numbers in the basin and hurricane numbers
    at landfall is taken to be linear and poisson distributed

\end{itemize}

This article proceeds as follows.
In section~\ref{assumptions} we discuss our choice of models and assumptions.
In section~\ref{models} we discuss the types of statistical models we will use, and the terminology
we will use to describe them.
In section~\ref{setup} we describe our notation and what we need from the SST forecasts.

In section~\ref{s2h-ln} we derive expressions for aspects of the predicted distribution of
hurricane numbers in the case where the relationship between
SST and hurricane numbers is linear and normally distributed.
In section~\ref{s2h-lp} we derive expressions for aspects of the predicted distribution of
hurricane numbers in the case where the relationship between
SST and hurricane numbers is linear and \emph{poisson} distributed.
In section~\ref{s2h-ep} we derive expressions for aspects of the predicted distribution of
hurricane numbers in the case where the relationship between
SST and hurricane numbers is \emph{exponential} and poisson distributed.

In section~\ref{b2l-lp} we derive expressions for aspects of the predicted distribution of landfalling
hurricane numbers when predicted as a function of the number of basin hurricane numbers,
in the case where the relationship between the two is linear and poisson distributed.


Finally in section~\ref{s2l2b} we discuss how to predict landfalling hurricane numbers from
basin numbers which are themselves predicted from SST by
combining the relationships derived in
sections~\ref{s2h-ln}, \ref{s2h-lp} and \ref{s2h-ep} with the relationships derived in
section~\ref{b2l-lp}.

\section{Comments on our choice of models}\label{assumptions}

This paper derives various mathematical expressions related to the prediction of hurricane numbers.
It is not, however, a discussion of which models are actually appropriate to use to predict hurricane
numbers. This latter question is a question we discuss at great length elsewhere: for
a discussion of what models can be used to predict SST see~\citet{j92} and~\citet{e20};
for a discussion of what models can be used to relate SST to landfalling hurricane numbers see~\citet{e04a};
for a discussion of what models can be used to relate SST to basin hurricane numbers see~\citet{e04b};
and for a discussion
of what models can be used to relate basin hurricane numbers to landfalling hurricane numbers see~\citet{e05}.

However, we now give a brief summary of the rationale behind our model choices, based on the results from
these studies:

\begin{itemize}

    \item we consider a normal distribution for SST because that seems to be the simplest
    reasonable model

    \item we consider a linear-normal relationship between SST and hurricane numbers
    because this is the simplest case analytically, even though the use of the normal
    distribution for hurricane numbers may not be reasonable if the number of hurricanes
    is small

    \item we consider a linear-poisson relationship between SST and hurricane numbers
    because this is the simplest case that has a reasonable distribution for hurricane
    numbers even in the situation in which the number of hurricanes is small. We are not
    concerned that use of a linear relationship could in principle lead to a negative
    value for the poisson parameter since this does not happen in practice with the data
    we are using.

    \item we consider an exponential-poisson relationship between SST and hurricane numbers
    because this has been used previously (e.g. by~\citet{elsners93}), and is the standard way that statisticians
    tend to use poisson regression. However, from the analysis we describe in~\citet{e04a} we conclude
    that it is not possible to tell from the data whether this model is better or worse
    than the linear-poisson model.

    \item we consider a linear-poisson relationship between basin hurricane numbers
    and landfalling hurricane numbers because it is simple, it does
    well in our own tests with real data (see~\citet{e05}) and it includes
    the simpler model that consists of just a constant proportion.


\end{itemize}

\section{Comments on notation and statistical models}\label{models}

Consider trying to build a statistical model that models some variables $y_i$ as a function of some other variables $x_i$.
One obvious place to start is standard linear regression, which can be written as:
\begin{equation}
y_i=\alpha+\beta x_i+\epsilon_i
\end{equation}
where the $\epsilon_i$ are taken to be IID and normally distributed with mean zero.
From now on we drop the subscripts $i$ for simplicity.

We can also write this model as:
\begin{equation}\label{ln2}
y \sim N (\alpha+\beta x, \sigma^2)
\end{equation}

In this paper we will call this the linear-normal model.

This model can be generalised in a number of ways.
One way would be to change the distribution from normal to something else.
We could do this either by specifying a different distribution for the noise
forcing $\epsilon$, or by specifying a different distribution for the response $y$. In general
these are not equivalent: for instance if we were to specify that the noise should
be poisson distributed then $y$ would have a poisson distribution shifted by
$\alpha+\beta x$, and, conversely, if we were to specify that $y$ should be
poisson distributed then the noise would have a poisson distribution shifted by
$-\alpha-\beta x$. In our case we think we have more idea about the distribution of $y$
than we do about the distribution of the noise: since $y$ is modelling hurricane numbers
we think that it will be close to poisson.
So the first generalisation we consider will be to replace equation~\ref{ln2} with:
\begin{equation}
y \sim \mbox{Po} (\mbox{rate}=\alpha+\beta x)
\end{equation}

In other words: we model $y$ as poisson distributed with mean $\alpha+\beta x$.
We will call this the \emph{linear-poisson} model.

We can also write this model as
\begin{equation}
y=\alpha+\beta x+\epsilon
\end{equation}
just as before, but we note that although it is still the case that $E(\epsilon)=0$, the distribution
of $\epsilon$ is now somewhat odd, since it is a poisson distribution but shifted
to have zero mean. As a result it is not common to write this model in this way (although there is
nothing wrong with it).

We could also write this model using conditional expectations:
\begin{equation}
 E(y|x)=\alpha+\beta x
\end{equation}
with the additional information that $y$ is poisson distributed.

We note that the linear poisson model does not often appear in statistics textbooks
because for certain values of $\alpha, \beta$ and $x$ the rate of the poisson could be
negative, which is impossible. We, however, take a practical approach: for our
data this is not a problem, and what happens far outside the domain covered by our
data is not relevant to us and can be ignored.

Having discussed how to change the distribution used in standard linear regression,
the second obvious generalisation would be to replace the linear function
with something non-linear, and a common way to do this is
to replace $\alpha+\beta x$ with $\mbox{exp}(\alpha+\beta x)$.
For those concerned about the problem with negative poisson rates described above, this solves
that problem. We can write this new model as:

\begin{equation}
y \sim \mbox{Po} (\mbox{rate}=\mbox{exp}(\alpha+\beta x))
\end{equation}

or

\begin{equation}
y=\mbox{exp}(\alpha+\beta x)+\epsilon
\end{equation}

or

\begin{equation}
 E(y|x)=\mbox{exp}(\alpha+\beta x)
\end{equation}
where $y$ is poisson distributed.

In fact, statisticians often write this model in yet another way, as:
\begin{equation}
\mbox{log} E(y|x)=\alpha+\beta x
\end{equation}

and call it the log-linear poisson regression model, although we find this nomenclature rather
unhelpful, and we prefer to call this model an exponential-poisson regression model.

One final comment is that one can discuss all the models given above in the context of a general
class of models known as generalised linear models (GLMs), which cover almost any possible distribution
for $y$ and any possible non-linear monotonic function of $\alpha+\beta x$.

\section{Setup and basic relations}\label{setup}

We will use the following notation:

\begin{itemize}
    \item $s$ is the SST
    \item $s$ is taken as normal, with mean $\mu_s$ and sd $\sigma_s$
    \item $n$ is the number of hurricanes
    \item $p()$ is used for probability densities
    \item $E()$ is used for expectations
    \item when we need to distinguish between the number of basin hurricanes and the number of landfalling hurricanes
    we write these as $n_b$ and $n_l$.
    \item the relationships we use to model hurricane numbers as a function of SST give us the parameters $\alpha, \beta, \sigma$
    \item the relationships we use to model landfalling hurricane numbers as a function of basin hurricane numbers
    give us the parameters $\alpha', \beta', \sigma'$
    \item the mean and variance of the number of hurricanes is written as $\mu_h$ and $\sigma_h^2$,
    with $\mu_b$ and $\sigma_b^2$ for basin hurricanes and $\mu_l$ and $\sigma_l^2$ for landfalling
\end{itemize}

\subsection{Calculating Means}

We will calculate the mean number of hurricanes using:

\begin{eqnarray}
 \mu_h&=&E(n)\\
      &=&\sum_{n=0}^{\infty} n p(n) \\
      &=&\sum_{n=0}^{\infty} n \int_{-\infty}^{\infty} p(n|s) p(s) ds \\
      &=&\int_{-\infty}^{\infty} \left(\sum_{n=0}^{\infty} n p(n|s)\right) p(s) ds\\
      &=&\int_{-\infty}^{\infty} E(n|s) p(s) ds
\end{eqnarray}

\subsection{Variances}

We will calculate the variance of the number of hurricanes using:

\begin{eqnarray}
\sigma_h^2     &=&E [(n-\mu_h)^2]\\
               &=&\sum_{n=0}^{\infty} (n-\mu_h)^2 p(n) \\
               &=&\sum_{n=0}^{\infty} (n-\mu_h)^2 \int_{-\infty}^{\infty} p(n|s) p(s) ds \\
               &=&\int_{-\infty}^{\infty} \left(\sum_{n=0}^{\infty} (n-\mu_h)^2 p(n|s)\right) p(s) ds\\
               &=&\int_{-\infty}^{\infty} E((n-\mu_h)^2|s) p(s) ds
\end{eqnarray}

In the non-linear cases we will calculate the variances using:
\begin{eqnarray}
sigma_h^2       &=&E [(n-\mu_h)^2]\\
                &=&E (n^2) - \mu_h^2
\end{eqnarray}

where

\begin{eqnarray}
 E(n^2)&=&\sum_{n=0}^{\infty} n^2 p(n) \\
     &=&\sum_{n=0}^{\infty} n^2 \int_{n=0}^{\infty} p(n|s) p(s) ds \\
     &=&\int_{n=0}^{\infty} \left(\sum_{n=0}^{\infty} n^2 p(n|s)\right) p(s) ds\\
     &=&\int_{n=0}^{\infty} E(n^2|s) p(s) ds
\end{eqnarray}

\subsection{SST forecasts}\label{sst}

We assume that we have some method for the prediction of SST that gives
us a normal distribution $N(\mu_s, \sigma_s^2)$.
The parameters $\mu_s$ and $\sigma_s$ will typically have some uncertainty associated with
them, which will depend on the prediction method being used. Some of the
expressions we derive below need estimates of this uncertainty in order to derive
estimates of the standard errors on our hurricane number predictions.

%
%
%
%
%
%
%
%
%
%

\section{The linear-normal model for the relationship between SST and hurricane numbers}\label{s2h-ln}

We now describe the first of our models for the relationship between SST and hurricane numbers,
which is the linear-normal model.
In this model we postulate a linear relation between SST $s$ and hurricane numbers $n$, where $n$ could
be either the number of hurricanes in the basin, or the number at landfall. We assume that the
distribution of hurricane numbers is normal. As discussed above, we include this model because it is
perhaps the simplest model one might consider. However, the assumption of normality probably doesn't
hold very well when the number of hurricanes is small: this is resolved in section~\ref{s2h-lp} by
replacing the normal distribution with the poisson.

The linear-normal model can be written as:

\begin{eqnarray}
n        &=&    \alpha+\beta s+\epsilon\label{ln1}\label{n1}\\
\epsilon &\sim& N(0,\sigma^2)
\end{eqnarray}

\subsection{The predicted mean}

Taking expectations of equation~\ref{ln1} (over all realisations of $\epsilon$ and $s$)
gives us a simple expression for the mean number of predicted hurricanes in this model:

\begin{equation}
 \mu_h=E(n)=\alpha+\beta E(s)=\alpha+\beta \mu_s\label{mu1}
\end{equation}

We see that the mean number of hurricanes is a linear function of the mean SST, and doesn't depend
on the variance of SST $\sigma_s^2$.

\subsection{The predicted variance}

We can also derive an expression for the variance of the number of hurricanes $\sigma_h^2$ fairly easily.

Combining equations~\ref{n1} and~\ref{mu1} we see that:

\begin{eqnarray}
n-\mu_h     &=&\beta(s-\mu_s)+\epsilon\\
(n-\mu_h)^2 &=&\beta^2(s-\mu_s)^2+2\beta(s-\mu_s)\epsilon+\epsilon^2\\
E((n-\mu_h)^2|s) &=&\beta^2(s-\mu_s)^2+\sigma^2\\\nonumber
\end{eqnarray}

where we take expecatations over all realisations of $\epsilon$, for fixed $s$.

This gives:

\begin{eqnarray}
 \sigma_h^2  &=&\int E((n-\mu_h)^2|s) p(s) ds\\
             &=&\int (\beta^2 (s-\mu_s)^2+\sigma^2)p(s) ds\\
             &=&\beta^2 \int (s-\mu_s)^2 p(s) ds+\int \sigma^2 p(s) ds\\
             &=&\beta^2 \sigma_s^2+\sigma^2
\end{eqnarray}

This expression is easy to understand: the variance in the number of predicted hurricanes comes
both from the variance in the SST prediction (scaled by $\beta$) and the variance around the
relationship between mean SST and the mean number of hurricanes. The variance of the predicted number
(b) the standard error on the SST prediction (which should be given by the SST prediction routine).

\subsection{Standard errors}

We now derive an approximate expression for the standard error on the prediction of the expected
number of hurricanes.

We already have:
\begin{equation}
\mu_h=\alpha+\beta \mu_s
\end{equation}

Now consider small errors in the parameters $\alpha$ and $\beta$ and a small error in the prediction of the mean
SST. We can understand what errors this leads to in our prediction of expected hurricane numbers just by
linearising:

\begin{equation}
dm=d \alpha+\mu d \beta + \beta d\mu
\end{equation}

Squaring this and taking expectations gives:

\begin{equation}
\mbox{var}(\mu_h)=\mbox{var}(\alpha_s)+\mu^2 \mbox{var}(\beta)+\beta^2 \mbox{var}(\mu_s)+2\mu_s \mbox{cov}(\alpha,\beta)
\end{equation}

This gives us an approximation to the standard error on the expected number of hurricanes in terms of the
standard errors on the parameters from the regression (which are given by most regression routines), and the
standard error on the SST prediction (which should be given by the SST prediction routine).

\subsubsection{Higher order terms}

By using Taylor expansions, we can derive higher order terms in the expression for the standard error.
Consider the function $\mu_h=f(\alpha,\beta,\mu_s)$. Expanding this function in a Taylor
series gives:

\begin{eqnarray}
 d\mu_h&=&\frac{\partial \mu_h}{\partial \alpha} d\alpha
         +\frac{\partial \mu_h}{\partial \beta} d\beta
         +\frac{\partial \mu_h}{\partial \mu_s} d\mu_s\\
       &&+\frac{1}{2} \frac{\partial^2 \mu_h}{\partial \alpha^2} d\alpha^2
         +\frac{1}{2} \frac{\partial^2 \mu_h}{\partial \beta^2} d\beta^2
         +\frac{1}{2} \frac{\partial^2 \mu_h}{\partial \mu_s^2} d\mu_s^2\\
       &&+\frac{\partial^2 \mu_h}{\partial \alpha \partial \beta} d\alpha d\beta
         +\frac{\partial^2 \mu_h}{\partial \beta \partial \mu_s} d\beta d\mu_s
         +\frac{\partial^2 \mu_h}{\partial \mu_s \partial \alpha} d\mu_s d\alpha+...\\
       &=& d\alpha
         + \mu_s d\beta
         + \beta d\mu_s\\
       &&+0 d\alpha^2
         +0 d\beta^2
         +0 d\mu_s^2\\
       &&+0 d\alpha d\beta
         +1 d\beta d\mu_s
         +0 d\mu_s d\alpha+0\\
       &=&d\alpha+ \mu_s d\beta+ \beta d\mu_s+d\beta d\mu_s
\end{eqnarray}

This is exact, since all subsequent terms are zero.
Squaring and taking expectations now gives:
\begin{eqnarray}
\mbox{var}(d\mu_h)&=&\mbox{var}(\alpha)+\mu_s^2 \mbox{var}(\beta)+2 \mu_s \mbox{cov}(\alpha,\beta)+\beta^2\mbox{var}(\mu_s)
\end{eqnarray}

and we find that our original expression is actually exact.

\subsection{Linear sensitivity}

We can also consider the linear sensitivity of our forecast to changes in the mean and the standard deviation
of our SST prediction. This can be useful to understand how a change in the forecast will create a change
in the hurricane prediction.

Again starting with
\begin{equation}
\mu_h=\alpha+\beta \mu_s
\end{equation}

if we differentiate wrt $\mu_s$ we get:

\begin{equation}
\frac{\partial \mu_h}{\partial \mu_s}=\beta
\end{equation}

and
\begin{equation}
\frac{\partial \mu_h}{\partial \sigma_s}=0
\end{equation}

and we again see that the predicted mean number of hurricanes is independent of the variance in the SST forecast.

\subsection{Summary}\label{summary-s2h-ln}

We now summarise the relations we have derived for the linear-normal model:

\begin{eqnarray}
 \mu_h      &=& \alpha+\beta \mu_s\\
 \sigma_h^2 &=&\beta^2 \sigma_s^2+\sigma^2\\
\mbox{var}(\mu_h)&=&\mbox{var}(\alpha)+\mu_s^2 \mbox{var}(\beta)+\beta^2 \mbox{var}(\mu_s)+2\mu_s \mbox{cov}(\alpha,\beta)\label{se1}\\
\frac{\partial \mu_h}{\partial \mu_s}&=&\beta\\
\frac{\partial \mu_h}{\partial \sigma_s}&=&0
\end{eqnarray}

We didn't actually use the fact that the SST was normally distributed to derive these relations: all we used were the values
    for the first two moments of the SST distribution. So all the above results hold for any SST distribution, given
    the first two moments.

\subsection{Alternative representation}

In practice, equation~\ref{se1} is difficult to evaluate because the right hand side contains two large positive
terms ($\mbox{var}(\alpha)$ and $\mu_s^2 \mbox{var}(\beta)$) and one large negative term ($2\mu_s \mbox{cov}(\alpha,\beta)$).
Rounding error can easily cause the result to be negative (when it should be positive), or at least very inaccurate.

We can avoid this problem by rewriting the original regression equation as:
\begin{equation}
n=\alpha+\beta(s-\overline{s})+\epsilon
\end{equation}
where $\overline{s}$ is the mean of the observed values of historical SST.
Compared with the original formulation the value of $\beta$ is the same but the value of $\alpha$ is now
different.

This then gives:
\begin{equation}
 \mu_h=\alpha+\beta (\mu_s-\overline{s})
\end{equation}

The expression for the variance, which doesn't depend on $\alpha$, doesn't change.

The derivation for the standard errors is based on:
\begin{equation}
 d\mu_h=d\alpha+ (\mu_s-\overline{s}) d\beta+\beta d\mu_s
\end{equation}

giving:

\begin{equation}
\mbox{var} (\mu_h)=\mbox{var}(\alpha)+(\mu_s-\overline{s})^2 \mbox{var}(\beta)+\beta^2 \mbox{var}(\mu_s)
\end{equation}

The $\mbox{cov}(\alpha,\beta)$ term drops out because it is zero in this alternative representation.

\section{The linear-poisson model for the relationship between SST and hurricane numbers}\label{s2h-lp}

We now derive relations for the second of our models for the relationship between
SST and hurricane numbers. This model is the same as the previous model, except that
we replace the normal distribution with a poisson distribution. Because the expression
for the mean number of hurricanes given the SST is still linear the results are rather similar.

We write this model as:
\begin{eqnarray}
n        &=&\alpha+\beta s+\epsilon\\
n &\sim& \mbox{Po} (\mbox{rate}=\alpha+\beta s)
\end{eqnarray}

Note that the variance of $n$ given $s$ is given by $v(s)=\alpha+\beta s$
whereas in the previous model the variance was constant.

\subsection{The predicted mean}

Just as before:

\begin{equation}
 \mu_h=E(n)=\alpha+\beta \mu_s\label{mean}
\end{equation}

\subsection{The predicted variance}

\begin{eqnarray}
n-\mu_h     &=&\beta(s-\mu_s)+\epsilon\\
(n-\mu_h)^2 &=&\beta^2(s-\mu_s)^2+2\beta(s-\mu_s)\epsilon+\epsilon^2\\
E((n-\mu_h)^2|s) &=&\beta^2(s-\mu_s)^2+v(s)
\end{eqnarray}

where $v(s)=\alpha+\beta s$ and the expectation in the final step is again over realisations of the noise but
not over realisations of the SST.
The final equation is now slightly different than before because of the dependence of the variance on $s$.
This has implications for the next step:

\begin{eqnarray}
 \sigma_h^2  &=&\int E((n-\mu_h)^2|s) p(s) ds\\
       &=&\int (\beta^2 (s-\mu_s)^2+v(s))p(s) ds\\
       &=&\beta^2 \int (s-\mu_s)^2 ds+\int v(s) p(s) ds\\
       &=&\beta^2 \int (s-\mu_s)^2 ds+\int (\alpha+\beta s) p(s) ds\\
       &=&\beta^2 \sigma_s^2+(\alpha+\beta \mu_s)\\
       &=&\beta^2 \sigma_s^2+\mu_h\label{var}
\end{eqnarray}

We see that the part of the uncertainty in the hurricane number prediction
that depends on the uncertainty in the SST-hurricane relationship
now becomes $\mu_h$, the predicted mean number of hurricanes.
In other words, the higher the mean number of hurricanes predicted, the greater the
uncertainty (in absolute terms).

Comparing the expression for the mean (equation~\ref{mean})
and the expression for the variance (equation~\ref{var}) we see that,
in general, they are not the same. We conclude that the predicted hurricane distribution, although a mixture
of poisson distributions, is not itself a poisson distribution.

\subsection{Standard errors}

The derivation for the standard errors is the same as for the linear-normal case:

\begin{equation}
\mu_h=\alpha+\beta \mu_s
\end{equation}

and so

\begin{equation}
d\mu_h=d \alpha+\mu_s d \beta + \beta d\mu_s
\end{equation}

and
\begin{equation}
\mbox{var}(\mu_h)=\mbox{var}(\alpha)+\mu_s^2 \mbox{var}(\beta)+\beta^2 \mbox{var}(\mu_s)+2\mu_s \mbox{cov}(\alpha,\beta)
\end{equation}

As before, this is in fact exact.

\subsection{Linear sensitivity}

This is also the same as for the linear-normal case:

\begin{equation}
\mu_h=\alpha+\beta \mu_s
\end{equation}

so
\begin{equation}
\frac{\partial \mu_h}{\partial \mu_s}=\beta
\end{equation}

and
\begin{equation}
\frac{\partial \mu_h}{\partial \sigma_s}=0
\end{equation}

\subsection{Summary}\label{summary-s2h-lp}

\begin{eqnarray}
 \mu_h &=& \alpha+\beta \mu_s\\
 \sigma_h^2 &=&\beta^2 \sigma_s^2+\alpha+\beta \mu_s\\
\mbox{var}(\mu_h)&=&\mbox{var}(\alpha)+\mu_s^2 \mbox{var}(\beta)+\beta^2 \mbox{var}(\mu_s)+2\mu_s \mbox{cov}(\alpha,\beta)\\
\frac{\partial \mu_h}{\partial \mu_s}&=&\beta\\
\frac{\partial \mu_h}{\partial \sigma_s}&=&0
\end{eqnarray}

Once again the fact that we assumed SST was normal was irrelevant...and again all we used about the SST was
the information about the first two moments.

\subsection{Alternative representation}

The alternative representation is almost the same as for the linear-normal case:

\begin{equation}
n=\alpha+\beta(s-\overline{s})+\epsilon
\end{equation}

This then gives:
\begin{equation}
 \mu_h=\alpha+\beta (\mu_s-\overline{s})
\end{equation}

The derivation for the standard errors is based on:
\begin{equation}
 d\mu_h=d\alpha+ (\mu_s-\overline{s}) d\beta+\beta d\mu_s
\end{equation}

giving:

\begin{equation}
\mbox{var} (\mu_h)=\mbox{var}(\alpha)+(\mu_s-\overline{s})^2 \mbox{var}(\beta)+\beta^2 \mbox{var}(\mu_s)+2(\mu_s-\overline{s})\mbox{cov}(\alpha,\beta)
\end{equation}

In this case the covariance term is not necessarily zero.

\section{The exponential-poisson model for the relationship between SST and hurricane numbers}\label{s2h-ep}

We now consider the third of our models for the relationship between SST and hurricane numbers,
in which the mean number of hurricanes (given the SST) is given by an exponential of a linear
function of SST, and the distribution of the number of hurricanes is poisson.

\begin{eqnarray}
n      &=&\mbox{exp}(\alpha+\beta s)+\epsilon\\
n &\sim&\mbox{Po}(\mbox{rate}=\mbox{exp}(\alpha+\beta s))
\end{eqnarray}

The variance depends on the SST, as for the linear-poisson model.

There are now big differences in the following analysis because of the non-linearity.

\subsection{The predicted mean}

Because of the non-linearity, evaluating the mean number of hurricanes is now a bit harder.
We actually have to do the integral, and so we actually do have to use the assumption that the SST
distribution is normal, rather than just needing the first two moments as in the previous two models.

\begin{eqnarray}
 \mu_h
     &=&\int \mbox{exp}(\alpha+\beta s)p(s) ds \\
     &=&\int \mbox{exp}(\alpha+\beta s) \frac{1}{\sqrt{2\pi} \sigma_s} \mbox{exp}\left( -\frac{(s-\mu_s)^2}{2\sigma^2}\right) ds \\
     &=&\mbox{exp}(\alpha) \frac{1}{\sqrt{2\pi} \sigma_s} \int \mbox{exp}\left(\beta s -\frac{(s-\mu_s)^2}{2\sigma^2}\right) ds \\
     &=&\mbox{exp}(\alpha) \frac{1}{\sqrt{2\pi} \sigma_s}
        \int \mbox{exp}\left[ -\frac{1}{2 \sigma_s^2}(-2 \sigma_s^2\beta s +(s-\mu_s)^2\right] ds \\
     &=&\mbox{exp}(\alpha) \frac{1}{\sqrt{2\pi} \sigma}
        \int \mbox{exp}\left[ -\frac{1}{2 \sigma_s^2}(s^2-2(\mu+\beta \sigma_s^2)s+\mu_s^2)\right] ds \\
     &=&\mbox{exp}(\alpha) \frac{1}{\sqrt{2\pi} \sigma_s}
        \int \mbox{exp}\left[ -\frac{1}{2 \sigma_s^2}(s-(\mu+\beta \sigma_s^2))^2 -(\mu+\beta \sigma_s^2)^2+\mu_s^2) \right]ds\\
     &=&\mbox{exp}(\alpha) \frac{1}{\sqrt{2\pi} \sigma_s}
        \int \mbox{exp}\left[ -\frac{1}{2 \sigma_s^2}((s-a)^2 +b) \right]ds\\
     &=&\mbox{exp}\left(\alpha-\frac{b}{2\sigma_s^2}\right) \frac{1}{\sqrt{2\pi} \sigma_s}
        \int \mbox{exp}\left[ -\frac{1}{2 \sigma_s^2}((s-a)^2) \right]ds\\
     &=&\mbox{exp}\left(\alpha-\frac{b}{2\sigma_s^2}\right) \frac{1}{\sqrt{2\pi} \sigma_s}
        \sqrt{2\pi} \sigma_s\\
     &=&\mbox{exp}\left(\alpha-\frac{b}{2\sigma_s^2}\right)\\
     &=&\mbox{exp}\left(\alpha+\beta (\mu_s+\beta \sigma_s^2/2)\right)
     \label{meanrate}
\end{eqnarray}

where we used the temporary variables:
\begin{eqnarray}
 a&=&\mu_s+\beta \sigma_s^2\label{a}\\
 b&=&\mu_s^2-a^2\\
  &=&\beta \sigma_s^2 (2\mu_s-\beta \sigma_s^2)\label{b}
\end{eqnarray}

\subsection{The predicted variance}

We can derive an expression for the variance in this case as follows:

\begin{eqnarray}
\sigma_h^2&=&E [(n-\mu_h)^2]\\
               &=&E (n^2) - \mu_h^2
\end{eqnarray}

$\mu_h^2$ we know already, so we just need to calculate $E (n^2)$.

\begin{eqnarray}
E n^2 &=& \int E(n^2|s) p(s) ds\\
      &=& \int E[\mbox{exp}(\alpha+\beta s)+\epsilon]^2 p(s) ds\\
      &=& \int [\mbox{exp}(\alpha+\beta s)^2+v(s)] p(s) ds\\
      &=& \int [\mbox{exp}(2\alpha+2\beta s)+v(s)] p(s) ds\\
      &=& I_1+I_2
\end{eqnarray}

\begin{eqnarray}
I_1&=&\int \mbox{exp}(2\alpha+2\beta s) p(s) ds\\
   &=&\mbox{exp}(2\alpha+2\beta (\mu+2\beta\sigma_s^2/2))\\
   &=&[\mbox{exp}(\alpha+\beta (\mu+\beta\sigma_s^2/2))]^2\mbox{exp}(\beta^2 \sigma_s^2)\\
   &=&\mu_h^2\mbox{exp}(\beta^2 \sigma_s^2)
\end{eqnarray}

\begin{eqnarray}
I_2&=&\int v(s) p(s) ds\\
   &=&\int \mbox{exp}(\alpha+\beta s) p(s) ds\\
   &=&\mu_h
\end{eqnarray}

so
\begin{eqnarray}
\sigma_h^2     &=&E (n^2)-\mu_h^2\\
               &=&\mu_h^2 \mbox{exp}(\beta^2 \sigma_s^2)+\mu_h-\mu_h^2\\
               &=&\mu_h^2(\mbox{exp}(\beta^2 \sigma_s^2)-1)+\mu_h\\
               &\approx&\mu_h^2\beta^2 \sigma_s^2+\mu_h
\end{eqnarray}

where the final approximation will be accurate if $\beta^2 \sigma_s^2 <<1$.

\subsection{Standard errors}

The derivation of expressions for standard errors follows the same logic as before, but
is slightly more complicated because of the non-linearity:

The mean is given by:
\begin{eqnarray}
\mu_h&=&e ^{\alpha+\beta(\mu_s+\beta\sigma_s^2/2)}\\
&=&e^x
\end{eqnarray}

where $x=\alpha+\beta (\mu_s+\beta \sigma_s^2/2$).

\begin{eqnarray}
d\mu_h&=&\frac{\partial \mu_h}{\partial \alpha} d\alpha
    +\frac{\partial \mu_h}{\partial \beta} d\beta
    +\frac{\partial \mu_h}{\partial \mu_s} d\mu_s
    +\frac{\partial \mu_h}{\partial \sigma_s} d\sigma_s\\
   &=&e^x (d\alpha+(\mu_s+\beta\sigma_s^2)d\beta+\beta d\mu_s+\beta^2 \sigma_s d\sigma_s)
\end{eqnarray}

which gives:
\begin{eqnarray}
 \mbox{var}(\mu_h)&=&e^{2x}
               [\mbox{var}(\alpha)
               +(\mu_s+\beta\sigma_s^2) \mbox{var}(\beta)
               +2(\mu_s+\beta \sigma_s^2)\mbox{cov}(\alpha,\beta)\\
            && +\beta \mbox{var}(\mu_s)
               +\beta^2 \sigma_s \mbox{var}(\sigma_s)
               +2 \beta^3 \sigma_s \mbox{cov}(\mu_s,\sigma_s)]
\end{eqnarray}

Unlike the linear cases this is now not exact, and there is in fact an infinite series of
higher order terms. Evaluating the second order terms would be a useful way to assess
the accuracy of the linear approximation.

\subsection{Linear sensitivity}

We can also calculate the linear sensitivity as follows:

\begin{eqnarray}
\frac{\partial \mu_h}{\partial \mu_s}&=&\frac{\partial}{\partial \mu_s} e^x \\
 &=& \beta e^x
\end{eqnarray}

This makes sense: the sensitivity to the mean SST is proportional to the mean, and
is proportional to $\beta$.

Also:
\begin{eqnarray}
\frac{\partial \mu_h}{\partial \sigma_s}&=&\frac{\partial}{\partial \mu_s} e^x\\
 &=& \beta^2 \sigma_s e^x
\end{eqnarray}

or
\begin{equation}
\frac{1}{\sigma_s} \frac{\partial \mu_h}{\partial \sigma_s}= \beta^2 e^x
\end{equation}

and we see that it is fractional changes in $\sigma_s$ that matter.
This is the only one of our three models where the standard deviation of the SST prediction
$\sigma_s$ has an impact on the mean hurricane numbers predicted. This arises because of the non-linearity:
more uncertainty on the SST prediction leads to higher expected numbers of hurricanes.

\subsection{Summary}\label{summary-s2h-ep}

\begin{eqnarray}
 \mu_h &=&\mbox{exp}\left(\alpha+\beta (\mu_s+\beta \sigma_s^2/2)\right)\\
 \sigma_h^2 &=&\mu_h^2(\mbox{exp}(\beta^2 \sigma_s^2)-1)+\mu_h\\
 \mbox{var}(\mu_h)&=&e^{2x}
               [\mbox{var}(\alpha)
               +(\mu_s+\beta\sigma_s^2) \mbox{var}(\beta)
               +2(\mu_s+\beta \sigma_s^2)\mbox{cov}(\alpha,\beta)\\
            && +\beta \mbox{var}(\mu_s)
               +\beta^2 \sigma_s \mbox{var}(\sigma_s)
               +2 \beta^3 \sigma_s \mbox{cov}(\mu_s,\sigma_s)]\\
\frac{\partial \mu_h}{\partial \mu_s}&=& \beta e^x\\
\frac{\partial \mu_h}{\partial \sigma_s}&=& \beta^2 \sigma_s e^x
\end{eqnarray}

\subsection{Alternative representation}

The alternative representation is also slightly different than before:

\begin{equation}
n=\mbox{exp}(\alpha+\beta(s-\overline{s}))+\epsilon
\end{equation}

This then gives:
\begin{equation}
 \mu_h=\mbox{exp}(\alpha+\beta(\mu_s-\overline{s}+\beta \sigma_s^2/2))
\end{equation}

The derivation for the standard errors is based on:
\begin{eqnarray}
d\mu_h&=&e^x (d\alpha+(\mu_s-\overline{s}+\beta\sigma^2)d\beta+\beta d\mu_s+\beta^2 \sigma_s d\sigma_s)
\end{eqnarray}

giving:
\begin{eqnarray}
 \mbox{var}(\mu_h)&=&e^{2x}
               [\mbox{var}(\alpha)
               +(\mu_s-\overline{s}+\beta\sigma_s^2) \mbox{var}(\beta)
               +2(\mu_s-\overline{s}+\beta \sigma_s^2)\mbox{cov}(\alpha,\beta)\\
            && +\beta \mbox{var}(\mu_s)
               +\beta^2 \sigma_s \mbox{var}(\sigma_s)
               +2 \beta^3 \sigma_s \mbox{cov}(\mu_s,\sigma_s)]
\end{eqnarray}

In this case the $\mbox{cov}(\alpha,\beta)$ term doesn't disappear, but will be much smaller than in the
original formulation.

\section{The linear-poisson model for the relationship between basin and landfalling hurricane numbers}\label{b2l-lp}

We now consider the relationship between the number of hurricanes in the basin and the number at
landfall.
We will use our model for this relationship later when we consider the possibility of predicting
landfalling hurricane numbers in a three step approach:
by first predicting SST, then predicting basin hurricane numbers
from SST, and finally predicting landfalling hurricane numbers from basin hurricane numbers.

The only model we consider in this case is a linear-poisson model for the number of landfalling hurricanes as a function
of the number of hurricanes in the basin:

\begin{eqnarray}
n_l      &=&\alpha'+\beta' n_b+\epsilon'\\
n_l &\sim&\mbox{Po}(\mbox{rate}=\alpha'+\beta' n_b)
\end{eqnarray}

Comparing with the previous linear-poisson model described in section~\ref{s2h-lp},
the SST $s$ has now become $n_b$.

There are only slight differences in the analysis compared with that model.

\subsection{The predicted mean}

\begin{equation}
 \mu_l=E(n)=\alpha'+\beta' \mu_b
\end{equation}

\subsection{The predicted variance}

\begin{eqnarray}
n-\mu_l     &=&\beta'(n_b-\mu_b)+\epsilon'\\
(n-\mu_l)^2 &=&\beta'^2(n_b-\mu_b)^2+2\beta'(n_b-\mu_b)e+\epsilon'^2\\
E((n-\mu_l)|n_b)^2 &=&\beta'^2(n_b-\mu_b)^2+v(n_b)\\
\end{eqnarray}

where $v(n_b)=\alpha'+\beta' n_b$.

\begin{eqnarray}
 \sigma_l^2  &=&\sum_{n=0}^{\infty} E((n-\mu_l)^2|n_b) p(n_b)\\
       &=&\sum_{n=0}^{\infty} (\beta'^2 (n_b-\mu_b)^2+v(n_b) )p(n_b)\\
       &=&\beta'^2 \sum_{n=0}^{\infty} (n_b-\mu)^2 +\sum_{n=0}^{\infty} v(n_b) p(n_b)\\
       &=&\beta'^2 \sigma_b^2+\alpha'+\beta' \mu_b\\
       &=&\beta'^2 \sigma_b^2+\mu_l
\end{eqnarray}


Once again, comparing the expression for the mean and the expression for the variance we see that,
in general, they are not the same. The predicted hurricane distribution, being a mixture
of poisson distributions, is not itself a poisson distribution.

\subsection{Standard errors}

\begin{equation}
\mu_l=\alpha'+\beta' \mu_b
\end{equation}
so
\begin{equation}
d\mu_l=d \alpha'+\mu_b d \beta' + \beta' d\mu_b
\end{equation}
and
\begin{equation}
\mbox{var}(\mu_l)=\mbox{var}(\alpha')+\mu_b^2 \mbox{var}(\beta')+\beta'^2 \mbox{var}(\mu_b)+2\mu_b \mbox{cov}(\alpha',\beta')
\end{equation}

As in the previous linear cases, this is exact.

\subsection{Linear sensitivity}

\begin{equation}
\mu_l=\alpha'+\beta' \mu_b
\end{equation}

so
\begin{equation}
\frac{\partial \mu_l}{\partial \mu_b}=\beta'
\end{equation}

It doesn't make sense to consider $\frac{\partial \mu_l}{\partial \sigma_b}$ since
$\sigma_b^2=\mu_b$.

\subsection{Summary}\label{summary-b2l-lp}

\begin{eqnarray}
 \mu_l &=& \alpha'+\beta' \mu_b\\
 \sigma_l^2 &=&\beta'^2 \sigma_b^2+\mu_l\\
\mbox{var}(\mu_l)&=&\mbox{var}(\alpha')+\mu_b^2 \mbox{var}(\beta')+\beta'^2 \mbox{var}(\mu_b)+2\mu_b \mbox{cov}(\alpha',\beta')\\
\frac{\partial \mu_l}{\partial \mu_b}&=&\beta'
\end{eqnarray}

\subsection{Alternative representation}

\begin{eqnarray}
 n_l& =    &\alpha'+\beta' (n_b-\overline{n_b})+\epsilon'\\
 n_l& \sim &\mbox{Po}(\mbox{rate}=\alpha'+\beta' (n_b-\overline{n_b}))
\end{eqnarray}

\begin{equation}
\mu_l=\alpha'+\beta' (\mu_b-\overline{n_b})
\end{equation}

\begin{equation}
d\mu_l=d\alpha'+(\mu_b-\overline{n_b}) d\beta'+\beta' d\mu_b
\end{equation}

\begin{equation}
 \mbox{var}(\mu_l)=\mbox{var}(\alpha')+(\mu_b-\overline{n_b})^2\mbox{var}(\beta')+\beta'^2\mbox{var}(\mu_b)+2(\mu_b-\overline{n_b})\mbox{cov}(\alpha',\beta')
\end{equation}

\section{Predicting landfalls from basin numbers, and basin numbers from SST}\label{s2l2b}

The models given in sections~\ref{s2h-ln}, \ref{s2h-lp} and~\ref{s2h-ep}
for relating SST to hurricane numbers (which we now take
as the number of hurricanes in the basin) can now be combined with the model given in
section~\ref{b2l-lp} that relates the numbers
of hurricanes in the basin to the number at landfall.


We consider two cases below, based on the linear-poisson and exponential-poisson models for the
number of hurricanes in the basin, and combined with the linear-poisson model for the number of hurricanes at landfall.

\subsection{Linear poisson model from SST to basin, linear-poisson model from basin to landfall}

\subsubsection{Mean and variance}

From section~\ref{summary-s2h-lp} we see that given an
SST distribution $N(\mu_s,\sigma_s^2)$ the mean and variance of the distribution of the number of basin hurricanes is:
\begin{eqnarray}
\mu_b     &=&\alpha+\beta \mu_s\\
\sigma^2_b&=&\beta^2\sigma_s^2+\mu_b
\end{eqnarray}

and from section~\ref{summary-b2l-lp} we see that
 given the mean and variance of the number of basin hurricanes the mean and variance of the number of landfalling
hurricanes is:
\begin{eqnarray}
\mu_l     &=&\alpha'+\beta' \mu_b\\
\sigma^2_l&=&\beta'^2\sigma_b^2+\mu_l
\end{eqnarray}

Putting these together in order to get from SST to landfalls in one step, we get:
\begin{eqnarray}
\mu_l     &=&\alpha'+\beta' \mu_b\\
          &=&\alpha'+\beta' (\alpha+\beta \mu_s)\\
          &=&\alpha'+\beta' \alpha+\beta' \beta \mu_s\\
\sigma^2_l&=&\beta'^2\sigma_b^2+\mu_l\\
          &=&\beta'^2(\beta^2\sigma_s^2+\mu_b)+\mu_l\\
          &=&\beta'^2 \beta^2\sigma_s^2+\beta'^2\mu_b+\mu_l
\end{eqnarray}

\subsubsection{Standard errors}

For the SST to basin hurricanes part we have:
\begin{equation}
\mbox{var}(\mu_b)=
\mbox{var}(\alpha)+\mu_s^2 \mbox{var}(\beta)+\beta^2 \mbox{var}(\mu_s)+2\mu_s \mbox{cov}(\alpha,\beta)
\end{equation}

and for the basin to landfall part we have:
\begin{equation}
\mbox{var}(\mu_l)=\mbox{var}(\alpha')+\mu_b^2 \mbox{var}(\beta')+\beta'^2 \mbox{var}(\mu_b)+2\mu_b \mbox{cov}(\alpha',\beta')
\end{equation}

and these two expressions can easily be combined to give a one-step expression for $\mbox{var}(\mu_l)$.

\subsection{Exponential poisson model from SST to basin, linear-poisson model from basin to landfall}

\subsubsection{Mean and variance}

From section~\ref{summary-s2h-ep} we see that
given SST distribution $N(\mu_s,\sigma_s^2)$ the mean and variance of the distribution of the number of basin hurricanes is:
\begin{eqnarray}
\mu_b     &=&\mbox{exp}(\alpha+\beta(\mu_s+\beta \sigma_s^2/2))\\
\sigma^2_b&=&\mu_b^2(\mbox{exp}(\beta^2 \sigma_s^2)-1)+\mu_b
\end{eqnarray}

and from section~\ref{summary-b2l-lp} we again see that
given the mean and variance of the number of basin hurricanes the mean and variance of the number of landfalling
hurricanes is:
\begin{eqnarray}
\mu_l     &=&\alpha'+\beta' \mu_b\\
\sigma^2_l&=&\beta'^2\sigma_b^2+\mu_l
\end{eqnarray}

Putting these together:
\begin{eqnarray}
\mu_l     &=&\alpha'+\beta' \mbox{exp}(\alpha+\beta(\mu_s+\beta \sigma_s^2/2))\\
\sigma^2_l&=&\beta'^2\mu_b^2(\mbox{exp}(\beta^2 \sigma_s^2)-1)+\beta'^2 \mu_h+\mu_l
\end{eqnarray}

\subsubsection{Standard errors}

For the SST to basin hurricanes part we have:
\begin{equation}
\mbox{var}(\mu_b)=e^{2x}
               [\mbox{var}(\alpha)
               +(\mu+\beta\sigma^2) \mbox{var}(\beta)
               +\beta \mbox{var}(\mu)
               +\beta^2 \sigma \mbox{var}(\sigma)+2(\mu+\beta \sigma^2)\mbox{cov}(\alpha,\beta)]
\end{equation}

and for the basin to landfall part we have:
\begin{equation}
\mbox{var}(\mu_l)=\mbox{var}(\alpha')+\mu_b^2 \mbox{var}(\beta')+\beta'^2 \mbox{var}(\mu_b)+2\mu_b \mbox{cov}(\alpha',\beta')
\end{equation}

and again these two expressions can easily be combined to give an expression for $\mbox{var}(\mu_l)$.

\clearpage
\bibliography{arxiv}

\begin{thebibliography}{6}
\providecommand{\natexlab}[1]{#1}
\providecommand{\url}[1]{\texttt{#1}}
\expandafter\ifx\csname urlstyle\endcsname\relax
  \providecommand{\doi}[1]{doi: #1}\else
  \providecommand{\doi}{doi: \begingroup \urlstyle{rm}\Url}\fi

\bibitem[Binter et~al.(2006{\natexlab{a}})Binter, Jewson, and Khare]{e05}
R~Binter, S~Jewson, and S~Khare.
\newblock {Statistical modelling of the relationship between hurricane numbers
  in the Atlantic Basin and at US landfall}.
\newblock \emph{RMS Internal Report E05}, 2006{\natexlab{a}}.

\bibitem[Binter et~al.(2006{\natexlab{b}})Binter, Jewson, and Khare]{e04a}
R~Binter, S~Jewson, and S~Khare.
\newblock {Statistical modelling of the relationship between Main Development
  Region Sea Surface Temperature and Atlantic Basin hurricane numbers}.
\newblock \emph{RMS Internal Report E04a}, 2006{\natexlab{b}}.

\bibitem[Binter et~al.(2006{\natexlab{c}})Binter, Jewson, and Khare]{e04b}
R~Binter, S~Jewson, and S~Khare.
\newblock {Statistical modelling of the relationship between Main Development
  Region Sea Surface Temperature and landfalling hurricane numbers}.
\newblock \emph{RMS Internal Report E04b}, 2006{\natexlab{c}}.

\bibitem[Elsner and Schmertmann(1993)]{elsners93}
J~Elsner and C~Schmertmann.
\newblock {Improving extended-range seasonal predictions of intense Atlantic
  hurricane activity}.
\newblock \emph{{Weather and Forecasting}}, 3:\penalty0 345--351, 1993.

\bibitem[Laepple et~al.(2006)Laepple, Jewson, Meagher, O'Shay, and Penzer]{e20}
T~Laepple, S~Jewson, J~Meagher, A~O'Shay, and J~Penzer.
\newblock {Five-year ahead prediction of Sea Surface Temperature in the
  Tropical Atlantic: a comparison of simple statistical methods}.
\newblock \emph{arXiv:physics/0701162}, 2006.

\bibitem[Meagher and Jewson(2006)]{j92}
J~Meagher and S~Jewson.
\newblock {Year ahead prediction of hurricane season SST in the tropical
  Atlantic}.
\newblock \emph{arxiv:physics/0606185}, 2006.

\end{thebibliography}

\end{document}